\documentstyle[11pt,paspconf,epsf]{article}

\def\etal{{\it et~al.\,}}
\def\mo{$M_\odot$}
\def\lo{$L_\odot\,$}
\def\mj{$M_{\rm J}\,$}

\def\mic{$\mu$m}
\def\sles{\lower2pt\hbox{$\buildrel {\scriptstyle <}
   \over {\scriptstyle\sim}$}}
\def\sgreat{\lower2pt\hbox{$\buildrel {\scriptstyle >}
   \over {\scriptstyle\sim}$}}

\begin{document}

\title{Advances in the Theory of Brown Dwarfs and Extrasolar Giant Planets}

\author{A. Burrows, D. Sudarsky, and C. Sharp}  
\affil{Department of Astronomy and Steward Observatory, 
                 University of Arizona, Tucson, AZ \ 85721}
\author{M. Marley}
\affil{Department of Astronomy, New Mexico State University, 
                 Box 30001/Dept. 4500, Las Cruces NM 88003}
\author{W.B. Hubbard and J.I. Lunine}
\affil{Lunar and Planetary Laboratory, University of Arizona,
                 Tucson, AZ \ 85721}
\author{T. Guillot}
\affil{Department of Meteorology, University of Reading, P.O. Box 239,
                 Whiteknights, Reading RG6 6AU, United Kingdom}
\author{D. Saumon}
\affil{Department of Physics and Astronomy, Vanderbilt University, Nashville, TN 37235}
\author{R. Freedman}
\affil{Space Physics Research Institute, NASA Ames Research Center, Moffett Field CA 94035}

\begin{abstract}

We have developed a new non--gray theory of the evolution, spectra, and colors of extrasolar
giant planets (EGPs) and brown dwarfs that reveals their exotic nature and uniqueness.
We have discovered that the fluxes of such objects for T$_{\rm eff}$s
from 1300 K to 100 K can be spectacularly higher in the near infrared bands than black body
values and that their infrared colors are quite blue.  As a consequence, EGPs and brown 
dwarfs reside in hitherto unoccupied realms of the H--R diagram and may be more easily found with current
and planned telescopes than previously imagined.

\end{abstract}


\section{Introduction}  \label{intro}

Doppler spectroscopy has now revealed about 20 objects
in the giant planet/brown dwarf regime, including companions to $\tau$ Boo, 51 Peg, $\upsilon$ And,
55 Cnc, $\rho$ CrB, 70 Vir, 16 Cyg, and 47 UMa (Butler \etal\ 1997; Cochran \etal\ 1997;
Marcy \& Butler 1996; Butler \& Marcy 1996; Mayor \& Queloz 1995;
Latham \etal\ 1989).  Furthermore,  the brown dwarf, Gl229B, has been discovered
(Oppenheimer \etal\ 1995; Nakajima \etal\ 1995; Matthews \etal\ 1996; Geballe \etal\ 1996).
Gl229B is a milestone because
it displays methane spectral features
and low surface fluxes that are unique to objects with effective temperatures (in
this case, T$_{\rm eff}$$\sim$950 K)
below the solar--metallicity main sequence edge.
In 1995 and 1996, we published a gray theory of the evolution of extrasolar 
giant planets (EGPs) with masses from 0.3 \mj
to 15 \mj, where~\mj denotes a Jupiter mass ($\sim$0.001 \mo)  
(Burrows \etal\ 1995; Saumon \etal\ 1996; Guillot \etal\ 1996).  

We (Burrows \etal\ 1997) have now developed a {\it non--gray} theory that  
encompasses the EGP/brown dwarf domains from 0.3 \mj to 70 \mj,
in aid of the direct searches
for substellar objects, be they ``planets'' or brown dwarfs, being planned
(TOPS and ExNPS reports; Leger \etal\ 1993).  
We have 
limited themselves to {\it solar--metallicity} objects in isolation and ignored the
effects of stellar insolation (Guillot \etal\ 1996).  In this communication, we 
summarize some of the results of this  
extensive new study, to which we refer the reader for details.

\section{Modeling Technique}

The opacities we employed are from extensions of the HITRAN database (Rothman \etal 1992, 1997),
the GEISA database (Husson \etal\ 1997), and theoretical calculations  
(Tyuterev \etal\ 1994; Goorvitch 1994; Tipping 1990; Wattson \& Rothman 1992; L. R. Brown, private communication).
For water, we used the new Partridge \& Schwenke H$_2$O database.
Our line list includes $1.9\times10^6$ lines
for CH$_4$ and CH$_3$D, $99,000$ lines for CO,  $11,400$ lines for NH$_3$,
$11,240$ lines for PH$_3$, and $179,000$ lines for H$_2$S.
Modeled continuum opacity sources include
$\rm H^-$
and $\rm H_2^-$ opacity
and collision--induced absorption (CIA)
of H$_2$ and helium (Borysow \& Frommhold 1990; Zheng \& Borysow 1995).  The latter is a direct function of pressure
and a major process in EGP/brown dwarf atmospheres.

To calculate atmosphere profiles and spectra, we used the k--coefficient
method (Goody \etal\ 1989; Lacis \& Oinas 1991), widely used in
planetary atmosphere modeling (see M. Marley, this volume).  This is not the ODF technique (Saxner \& Gustafsson 1984)
and gives excellent
agreement with full line--by--line computations of atmospheric transmission
(Grossman \& Grant 1992,1994a,1994b).

Chemical equilibrium calculations were performed with the ATLAS code and
data from Kurucz (1970). The Kurucz reaction constants are inaccurate at
low temperatures, but the NH$_3$ $\rightarrow$ N$_2$ and CH$_4$ $\rightarrow$ CO
conversions that occur in EGPs and brown dwarfs do so in regions of $T-P$ space for which the
Kurucz reaction constants are accurate.
Condensation of NH$_3$, H$_2$O, Fe, and
MgSiO$_3$ was included using data from various sources, including Eisenberg \& Kauzmann (1969), the
Handbook of Chemistry and Physics (1993), and Lange's Handbook of Chemistry
(1979). Following Fegley \& Lodders (1994, 1996), we assumed that Al, Ca,
Ti and V were removed either by condensation or were dissolved in silicate
grains at about the MgSiO$_3$ condensation temperature.

\section{Evolutionary Models}

In Burrows \etal (1995) and Saumon \etal (1996), we published cooling curves for
EGPs and small brown dwarfs that were based upon our then--current atmosphere models.
In Burrows \etal\ (1997), we have updated the H$_2$ CIA, H$_2$O, and CH$_4$ opacities and the radiative transfer algorithm,
as well as the chemical equilibrium calculations. 
Consequently, the evolutionary tracks have changed, but generally by no more than 10\%
in luminosity at any given time, for any given mass.  

Figure 1 depicts the updated luminosity versus time plots for objects from Saturn's mass (0.3 \mj) to 0.2 \mo.
It covers three orders of magnitude in mass and encapsulates the characteristics of the
entire family of substellar objects and the transition to stars.

\begin{figure}
\vspace{3.50in}
\caption{
Evolution of the luminosity (in L${_\odot}$) of solar--metallicity M dwarfs and substellar objects
versus time (in years) after formation.
The stars, ``brown dwarfs'' and ``planets'' are shown as solid, dashed, and dot--dashed
curves, respectively.
In this figure, we arbitrarily designate as ``brown dwarfs'' those objects that burn deuterium,
while we designate those that do not as ``planets.''
The masses in M${_\odot}$ label most of the curves, with the lowest three
corresponding to the mass of Saturn, half the mass of Jupiter, and the mass of Jupiter.
}
\label{fig-1}
\hbox to\hsize{\hfill\includegraphics{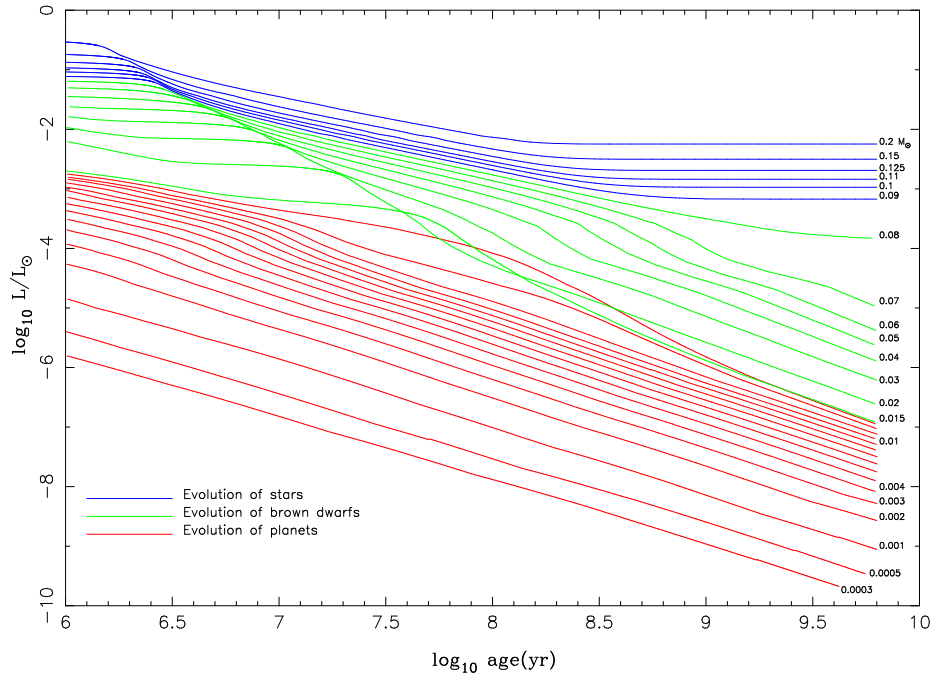}\kern+0in\hfill}
\end{figure}

Tables 1--3 summarize the evolution of various structural parameters versus time 
for EGP masses of  4 \mj, 15 \mj, and 20 \mj.

\begin{table}
\caption{Evolution of planet/brown dwarf of mass = 4 Jupiter masses}
\begin{center}\scriptsize
\begin{tabular}{crrrrrr}
$\log  t$ (Gyr)  & $T_{\rm eff}$ (K)  & $\log L/L_{\odot}$  & $R$ ($10^9$ cm)
& $\log T_c$ (K)  & $\log \rho_c$ (g cm$^{-3})$  &  $L_{\rm nuclear}/L$   \\
\tableline
$ -2.998 $&$  1652.0 $&$  -3.666 $&$  12.46$&$  5.132 $&$ 0.561 $&$0.000 $ \\
$ -2.794 $&$  1510.0 $&$  -3.877 $&$  11.72$&$  5.127 $&$ 0.627 $&$0.000 $ \\
$ -2.592 $&$  1353.0 $&$  -4.112 $&$  11.12$&$  5.118 $&$ 0.685 $&$0.000 $ \\
$ -2.385 $&$  1237.0 $&$  -4.310 $&$  10.59$&$  5.104 $&$ 0.740 $&$0.000 $ \\
$ -2.183 $&$  1112.0 $&$  -4.533 $&$  10.14$&$  5.085 $&$ 0.792 $&$0.000 $ \\
$ -1.976 $&$   981.9 $&$  -4.783 $&$   9.75$&$  5.061 $&$ 0.840 $&$0.000 $ \\
$ -1.768 $&$   860.4 $&$  -5.042 $&$   9.44$&$  5.034 $&$ 0.881 $&$0.000 $ \\
$ -1.558 $&$   748.9 $&$  -5.308 $&$   9.17$&$  5.004 $&$ 0.918 $&$0.000 $ \\
$ -1.346 $&$   648.5 $&$  -5.579 $&$   8.94$&$  4.974 $&$ 0.949 $&$0.000 $ \\
$ -1.136 $&$   562.7 $&$  -5.845 $&$   8.75$&$  4.944 $&$ 0.977 $&$0.000 $ \\
$ -0.927 $&$   490.3 $&$  -6.100 $&$   8.59$&$  4.914 $&$ 1.000 $&$0.000 $ \\
$ -0.715 $&$   427.3 $&$  -6.354 $&$   8.44$&$  4.884 $&$ 1.022 $&$0.000 $ \\
$ -0.503 $&$   373.3 $&$  -6.603 $&$   8.30$&$  4.852 $&$ 1.041 $&$0.000 $ \\
$ -0.303 $&$   328.7 $&$  -6.836 $&$   8.19$&$  4.822 $&$ 1.058 $&$0.000 $ \\
$ -0.100 $&$   288.5 $&$  -7.074 $&$   8.08$&$  4.790 $&$ 1.074 $&$0.000 $ \\
$  0.103 $&$   253.0 $&$  -7.313 $&$   7.98$&$  4.757 $&$ 1.088 $&$0.000 $ \\
$  0.306 $&$   222.2 $&$  -7.548 $&$   7.89$&$  4.724 $&$ 1.102 $&$0.000 $ \\
$  0.507 $&$   196.8 $&$  -7.769 $&$   7.81$&$  4.691 $&$ 1.114 $&$0.000 $ \\
$  0.709 $&$   176.2 $&$  -7.969 $&$   7.73$&$  4.654 $&$ 1.126 $&$0.000 $ \\
$  0.912 $&$   158.2 $&$  -8.166 $&$   7.65$&$  4.613 $&$ 1.138 $&$0.000 $ \\
\end{tabular}
\end{center}
\end{table}

\begin{table}
\caption{Evolution of planet/brown dwarf of mass = 15 Jupiter masses}
\begin{center}\scriptsize
\begin{tabular}{crrrrrr}
$\log  t$ (Gyr)  & $T_{\rm eff}$ (K)  & $\log L/L_{\odot}$  & $R$ ($10^9$ cm)
& $\log T_c$ (K)  & $\log \rho_c$ (g cm$^{-3})$  &  $L_{\rm nuclear}/L$   \\
\tableline
$ -2.993 $&$  2522.0 $&$  -2.701 $&$  16.24$&$  5.609 $&$ 0.840 $&$0.020 $ \\
$ -2.777 $&$  2484.0 $&$  -2.793 $&$  15.07$&$  5.624 $&$ 0.932 $&$0.054 $ \\
$ -2.565 $&$  2431.0 $&$  -2.902 $&$  13.88$&$  5.638 $&$ 1.036 $&$0.155 $ \\
$ -2.348 $&$  2369.0 $&$  -3.017 $&$  12.80$&$  5.648 $&$ 1.139 $&$0.385 $ \\
$ -2.129 $&$  2311.0 $&$  -3.114 $&$  12.04$&$  5.653 $&$ 1.217 $&$0.680 $ \\
$ -1.916 $&$  2264.0 $&$  -3.181 $&$  11.61$&$  5.654 $&$ 1.264 $&$0.851 $ \\
$ -1.713 $&$  2220.0 $&$  -3.239 $&$  11.30$&$  5.653 $&$ 1.299 $&$0.881 $ \\
$ -1.512 $&$  2143.0 $&$  -3.333 $&$  10.87$&$  5.651 $&$ 1.349 $&$0.843 $ \\
$ -1.310 $&$  1956.0 $&$  -3.545 $&$  10.23$&$  5.644 $&$ 1.428 $&$0.744 $ \\
$ -1.109 $&$  1621.0 $&$  -3.940 $&$   9.45$&$  5.623 $&$ 1.533 $&$0.544 $ \\
$ -0.904 $&$  1325.0 $&$  -4.354 $&$   8.77$&$  5.586 $&$ 1.630 $&$0.262 $ \\
$ -0.704 $&$  1095.0 $&$  -4.739 $&$   8.25$&$  5.536 $&$ 1.710 $&$0.108 $ \\
$ -0.503 $&$   890.3 $&$  -5.136 $&$   7.91$&$  5.487 $&$ 1.765 $&$0.054 $ \\
$ -0.301 $&$   740.3 $&$  -5.481 $&$   7.68$&$  5.442 $&$ 1.801 $&$0.030 $ \\
$ -0.099 $&$   628.2 $&$  -5.785 $&$   7.52$&$  5.401 $&$ 1.829 $&$0.017 $ \\
$  0.115 $&$   526.9 $&$  -6.106 $&$   7.39$&$  5.361 $&$ 1.851 $&$0.010 $ \\
$  0.328 $&$   450.7 $&$  -6.390 $&$   7.28$&$  5.325 $&$ 1.869 $&$0.006 $ \\
$  0.536 $&$   390.6 $&$  -6.649 $&$   7.20$&$  5.290 $&$ 1.884 $&$0.004 $ \\
$  0.740 $&$   339.9 $&$  -6.900 $&$   7.12$&$  5.255 $&$ 1.897 $&$0.002 $ \\
$  0.946 $&$   295.7 $&$  -7.150 $&$   7.05$&$  5.221 $&$ 1.909 $&$0.001 $ \\
\end{tabular}
\end{center}
\end{table}

\begin{table}
\caption{Evolution of planet/brown dwarf of mass = 20 Jupiter masses}
\begin{center}\scriptsize
\begin{tabular}{crrrrrr}
$\log  t$ (Gyr)  & $T_{\rm eff}$ (K)  & $\log L/L_{\odot}$  & $R$ ($10^9$ cm)
& $\log T_c$ (K)  & $\log \rho_c$ (g cm$^{-3})$  &  $L_{\rm nuclear}/L$   \\
\tableline
$ -2.988 $&$  2674.0 $&$  -2.251 $&$  24.27$&$  5.619 $&$ 0.489 $&$0.005 $ \\
$ -2.779 $&$  2668.0 $&$  -2.394 $&$  20.68$&$  5.664 $&$ 0.678 $&$0.052 $ \\
$ -2.563 $&$  2642.0 $&$  -2.532 $&$  18.00$&$  5.701 $&$ 0.848 $&$0.392 $ \\
$ -2.344 $&$  2627.0 $&$  -2.597 $&$  16.88$&$  5.718 $&$ 0.928 $&$0.877 $ \\
$ -2.137 $&$  2620.0 $&$  -2.620 $&$  16.52$&$  5.723 $&$ 0.955 $&$0.933 $ \\
$ -1.933 $&$  2609.0 $&$  -2.657 $&$  15.97$&$  5.731 $&$ 0.998 $&$0.895 $ \\
$ -1.733 $&$  2564.0 $&$  -2.770 $&$  14.53$&$  5.752 $&$ 1.119 $&$0.660 $ \\
$ -1.530 $&$  2313.0 $&$  -3.201 $&$  10.88$&$  5.785 $&$ 1.493 $&$0.000 $ \\
$ -1.328 $&$  1871.0 $&$  -3.709 $&$   9.25$&$  5.762 $&$ 1.704 $&$0.000 $ \\
$ -1.123 $&$  1525.0 $&$  -4.127 $&$   8.62$&$  5.733 $&$ 1.797 $&$0.000 $ \\
$ -0.923 $&$  1325.0 $&$  -4.412 $&$   8.21$&$  5.701 $&$ 1.860 $&$0.000 $ \\
$ -0.721 $&$  1168.0 $&$  -4.668 $&$   7.88$&$  5.665 $&$ 1.913 $&$0.000 $ \\
$ -0.509 $&$  1002.0 $&$  -4.965 $&$   7.61$&$  5.624 $&$ 1.958 $&$0.000 $ \\
$ -0.307 $&$   856.9 $&$  -5.258 $&$   7.41$&$  5.585 $&$ 1.992 $&$0.000 $ \\
$ -0.097 $&$   734.7 $&$  -5.545 $&$   7.25$&$  5.546 $&$ 2.020 $&$0.000 $ \\
$  0.118 $&$   621.0 $&$  -5.853 $&$   7.12$&$  5.506 $&$ 2.043 $&$0.000 $ \\
$  0.331 $&$   527.6 $&$  -6.148 $&$   7.02$&$  5.471 $&$ 2.061 $&$0.000 $ \\
$  0.539 $&$   456.9 $&$  -6.408 $&$   6.94$&$  5.437 $&$ 2.076 $&$0.000 $ \\
$  0.740 $&$   400.2 $&$  -6.648 $&$   6.87$&$  5.405 $&$ 2.089 $&$0.000 $ \\
$  0.949 $&$   348.8 $&$  -6.895 $&$   6.80$&$  5.371 $&$ 2.101 $&$0.000 $ \\
\end{tabular}
\end{center}
\end{table}

The early plateaux in Figure 1 between 10$^6$ years and 10$^8$ years are due to deuterium burning, where 
the initial deuterium mass fraction was taken to be 2$\times$10$^{-5}$.  Deuterium burning occurs earlier,
is quicker, and is at higher luminosity for the more massive models, but can take as long
as 10$^{8}$ years for a 15 \mj object.  The mass below which less than 50\% of the ``primordial''
deuterium is burnt is $\sim$13 \mj (Burrows \etal\ 1995).  On this figure, we have arbitrarily
classed as ``planets'' those objects that do not burn deuterium and as ``brown dwarfs'' those that do burn deuterium,
but not light hydrogen.  While this distinction is physically motivated, we do not 
advocate abandoning the definition based on origin.  Nevertheless, the separation 
into M dwarfs, ``brown dwarfs'', and giant ``planets'' is useful for parsing by eye the information in the figure.

In Figure 1, the bumps between 10$^{-4}$ \lo and 10$^{-3}$ \lo and between 10$^{8}$ and 10$^{9}$ years,
seen on the cooling curves of objects from 0.03 \mo to 0.08 \mo, are due to silicate and iron grain formation.   These effects,
first pointed out by Lunine \etal\ (1989), occur for T$_{\rm eff}$s between 2500 K and 1300 K.
Since grain and cloud models are problematic, there still remains much to learn concerning
their role and how to model them (see F. Allard, this volume). 

Bolometric luminosity and age can be used to yield mass and radius.   Effective temperature
and mass can provide age and luminosity.  Fits to the UKIRT spectrum of Gl229B (Marley \etal\ 1996; 
Allard \etal\ 1996; Tsuji \etal\ 1996) 
give T$_{\rm eff}$$\sim$900--1000 K and g$\sim$10$^{5\pm0.5}$ cm s$^{-2}$.
One obtains a mass between 15 \mj and 60 \mj, with a best value near 35 \mj, and an age
between 10$^{8.5}$ and 10$^{9.5}$ years.  The wide range in inferred Gl229B parameters is a direct consequence 
of the weakness of our current constraints on gravity.

\subsection{Metallicity Dependence}

Recently, we have completed a set of evolutionary models that explore the characteristics of the edge of the
main sequence as a function of metallicity and atmosphere model (Saumon \etal\ 1997, in preparation). 
As a part of this effort, we generated
10$^{10}$--year isochrones for models with metallicities of solar, 0.1$\times$solar, 0.01$\times$solar,
0.001$\times$solar, and zero (Saumon \etal\ 1994), employing atmosphere 
models of our own manufacture, as well as those from  
Alexander (1996) and Allard \& Hauschildt (1995). 
Figure 2 depicts the nature of the transition from brown dwarfs to M dwarfs for this range of metallicities. 

\begin{figure}
\vspace{3.50in}
\caption{
Luminosity (in L${_\odot}$) versus mass (in solar masses) for a variety of atmosphere models and metallicities, at an age of
10$^{10}$ years.  We performed all of the evolutionary calculations using our own codes and approaches, but employed 
the atmosphere models of Allard \& Hauschildt
(1995), Alexander (1996), Burrows \etal\ (1993), and Saumon \etal\ (1994). 
Systematically, the luminosity of an M dwarf is higher, and that of a brown dwarf is lower, for lower metallicities. 
}
\label{fig-2}
\hbox to\hsize{\hfill\includegraphics{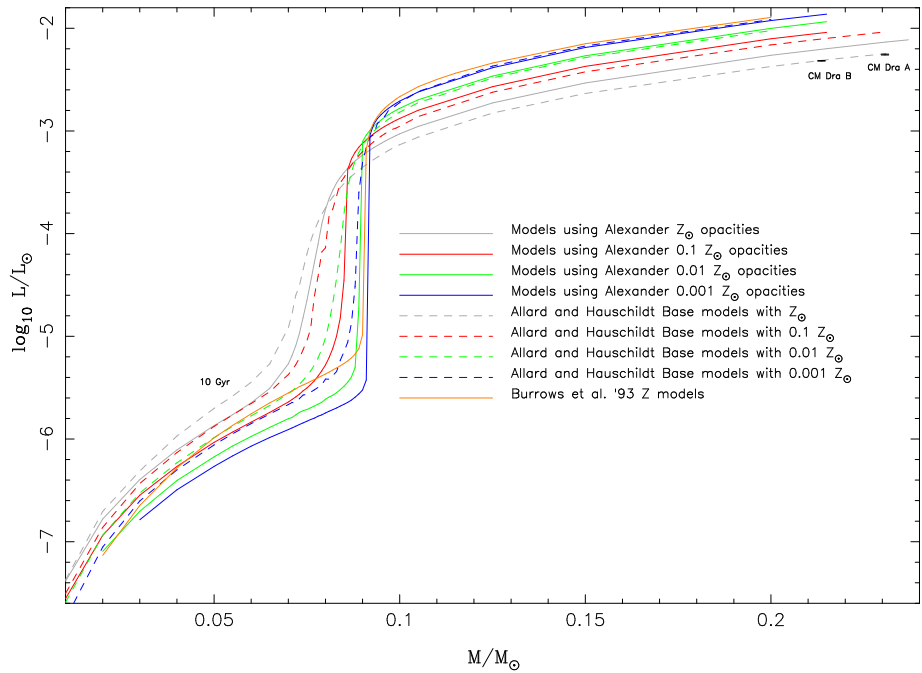}\kern+0in\hfill}
\end{figure}

We display these numerical data to caution the reader that the  metallicity dependence
is systematic and non--trivial.  
Figure 2 demonstrates clearly that the 
mass/luminosity slope at the boundary is a strong function of log$_{10}$Z.
In particular, the metallicity dependence will be a factor in the
proper conversion of luminosity functions into mass functions for objects in the field, in open clusters, and in globular
clusters.

\section{EGP and Brown Dwarf Spectra}

There are a few major aspects of EGP/brown dwarf atmospheres that bear listing and that uniquely
characterize them.  
Below T$_{\rm eff}$s of 1200 K, the dominant equilibrium carbon molecule is CH$_4$, not CO, 
and below 600 K the dominant nitrogen molecule is NH$_3$, not N$_2$.
The major opacity sources are H$_2$, H$_2$O, CH$_4$, and NH$_3$.
For T$_{\rm eff}$s below $\sim$400 K, water clouds form at or above the photosphere
and for T$_{\rm eff}$s below 200 K, ammonia clouds form ({\it viz.,} Jupiter).  Collision--induced absorption
of H$_2$ partially suppresses emissions longward of $\sim$10 \mic.  The holes in the opacity
spectrum of H$_2$O that define the classic telluric IR bands also regulate much of the emission from
EGP/brown dwarfs in the near infrared.  Importantly, the windows in H$_2$O and the suppression by H$_2$ conspire to
force flux to the blue for a given T$_{\rm eff}$.   
The upshot is an exotic spectrum enhanced relative to the black body value
in the $J$ and $H$ bands ($\sim$1.2 \mic\ and $\sim$1.6 \mic, respectively) by as much as {\it two} to {\it ten} orders of magnitude, 
depending upon T$_{\rm eff}$.  
In addition, as T$_{\rm eff}$ decreases below $\sim$1000 K, the flux in the $M$ band ($\sim$5 \mic)
is progressively enhanced relative to the black body value. 
While at 1000 K there is no enhancement, at 200 K it is near 10$^5$.   Hence, the $J$, $H$, and $M$ bands 
are the premier bands in which to search for cold substellar objects.    
Eventhough $K$ band ($\sim$2.2 \mic) fluxes are generally higher
than black body values,  H$_2$ and CH$_4$ absorption features in the $K$ band decrease its importance
{\it relative} to $J$ and $H$.  As a consequence of the increase of atmospheric 
pressure with decreasing T$_{\rm eff}$, the anomalously blue $J-K$ and $H-K$
colors get {\it bluer},
not redder.

Note that the presence or absence of clouds strongly affects the reflection albedos of EGPs and brown
dwarfs.  In particular, when there are clouds at or above the photosphere, the albedo in the optical
is high.  Conversely, when clouds are absent, the albedo in the mostly absorbing atmosphere
is low.  Recall, however, that in Burrows \etal\ (1997) we ignored reflection effects.

Figure 3 depicts the object's surface flux versus wavelength for representative T$_{\rm eff}$s from 130 K to
1000 K at a gravity of $3.0\times10^4$ cm s$^{-2}$.

\begin{figure}
\vspace{3.50in}
\caption{
Surface flux (in erg cm$^{-2}$ sec$^{-1}$ Hz$^{-1}$) versus wavelength (in microns)
from 1 \mic\ to 10 \mic\ for T$_{\rm eff}$s of 130, 200, 300, 500, 600, 700, and 1000 K,
at a surface gravity of $3.0\times10^4$ cm s$^{-2}$.  Shown are the positions of the $J$,
$H$, $K$, and $M$ bands and various molecular absorption features.  
}
\label{fig-3}
\hbox to\hsize{\hfill\includegraphics{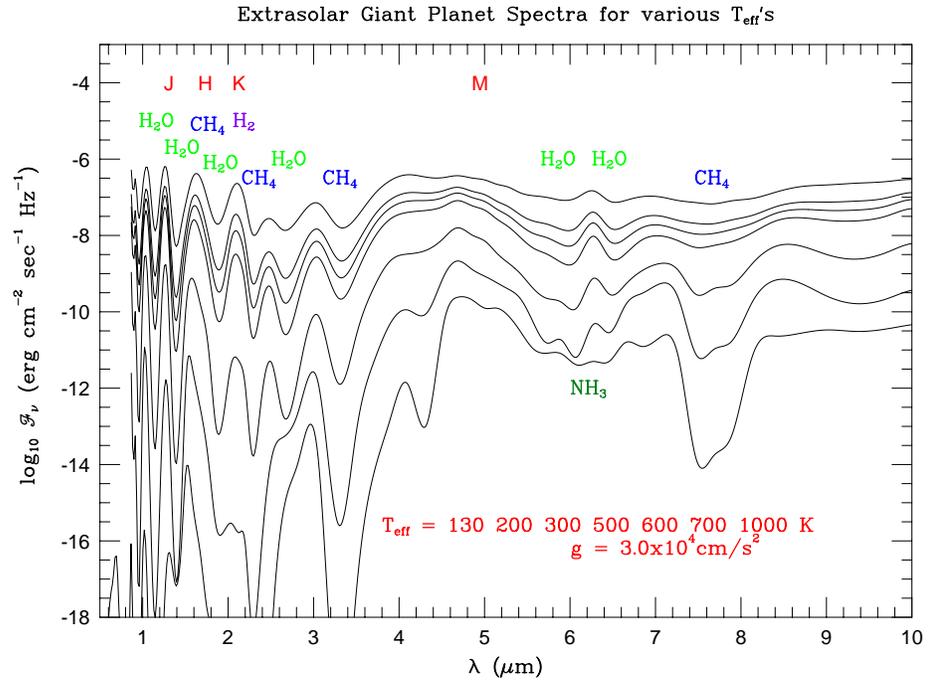}\kern+4in\hfill}
\end{figure}

The corresponding masses range 
from $\sim$13 \mj to $\sim$16 \mj and the corresponding ages range from 0.25 Gyrs to 7 Gyrs.  
Superposed on Figure 3 are the positions of various prominent molecular bands and the $J$, $H$,
$K$, and $M$ bands.  As is clear from the figure, H$_2$O defines much of the spectrum, but CH$_4$ and 
H$_2$ modify it in useful ways.  CH$_4$ absorption features near 1.65 \mic, 2.2 \mic, and 3.3 \mic\
are particularly relevant, the latter two decreasing the $K$ and $L^{\prime}$ ($\sim$3.5 \mic) band fluxes, respectively.  
NH$_3$ near 6 \mic\ becomes important below 250 K and the CH$_4$
feature around 7.8 \mic\ deepens with decreasing  T$_{\rm eff}$.   However, it should be noted
that in Jupiter the 7.8 \mic\ absorption feature is inverted into a stratospheric emission feature.
Since a stratosphere requires UV flux from the primary or another energy deposition mechanism, 
our models do not address this possibility.
In addition, we find that H$_2$O and NH$_3$ features near 6 \mic\ make the band from 5.5 to 7 \mic\
less useful for searching for brown dwarfs and EGPs.   

Figure 4 depicts spectra between 1 \mic\ and 40 \mic\ at a detector 
10 parsecs away from objects with age 1 Gyr and masses
from 1 \mj through 40 \mj.
Superposed are the corresponding black body curves and the putative sensitivities
for the three NICMOS cameras (Thompson 1992), ISO (Benvenuti \etal\ 1994), SIRTF (Erickson \& Werner 1992),
and Gemini/SOFIA (Mountain \etal\ 1992; Erickson 1992).
Figure 4 demonstrates how unlike a black body an EGP spectrum
is. 
Note on Figure 4 the H$_2$--induced suppression at long wavelengths and the enhancement at shorter
wavelengths.  For example, the enhancement at 5 \mic\ for a 1 Gyr old, 1 \mj$\!$ extrasolar planet is by four o
rders
of magnitude.

\begin{figure}
\vspace{3.50in}
\caption{
The flux (in $\mu$Janskys) at 10 parsecs versus wavelength (in microns) from 1 \mic\ to
40 \mic\ for 1, 5, 10, 20, 30, and 40 \mj models at 1 Gyr.   Superposed for comparison
are the corresponding black body curves (dashed) and the 
putative sensitivities of the three NICMOS cameras, ISO,
Gemini/SOFIA, and SIRTF.  NICMOS is denote with large black dots, ISO with thin, dark lines,
Gemini/SOFIA with thin, light lines, and SIRTF with thicker, dark lines.
At all wavelengths, SIRTF's projected sensitivity is greater than ISO's.
SOFIA's sensitivity overlaps with that of ISO around 10 \mic.  For other wavelength intervals, the order
of sensitivity is SIRTF $>$ Gemini/SOFIA $>$ ISO, where $>$ means ``is more sensitive than.''
Note the suppression relative to the black body values at the longer
wavelengths.
}
\label{fig-4}
\hbox to\hsize{\hfill\includegraphics{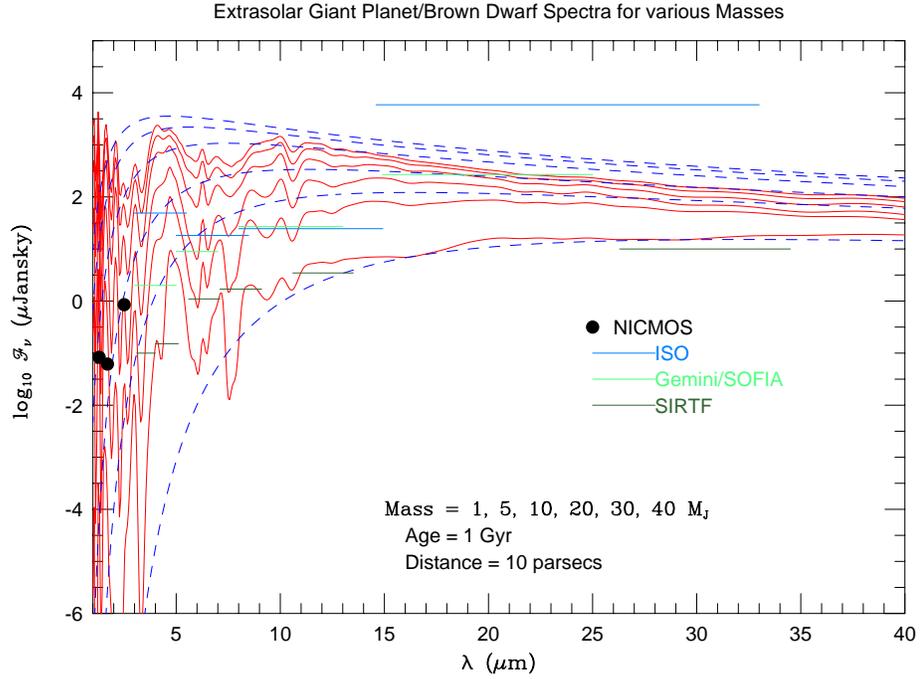}\kern+4in\hfill}
\end{figure}

Comparison with the sensitivities reveals that the range for detection by SIRTF at 5 \mic\ 
of a 1 Gyr old, 1 \mj object in isolation is near 100 parsecs.   The range for NICMOS in $H$
for a 1 Gyr old, 5 \mj object is approximately 300 parsecs, while for a coeval 40 \mj object
it is near 1000 parsecs.  These are dramatic numbers that serve to illustrate both the
promise of the new detectors and the enhancements we theoretically predict.

Figures 5--7 portray the evolution from 0.1 Gyr to 5 Gyr of the spectra from 1 \mic\ to 10 \mic\ 
of objects with masses of 1, 5, and 25 \mj.  
The higher curves are for the younger ages.
These cooling curves summarize EGP/brown dwarf
spectra and their evolution. 
Figure 7 suggests that SIRTF will be able to see at 5 \mic\ a 5 Gyr old, 25 \mj object in isolation out to
$\sim$500 parsecs and that NICMOS will be able to see at $J$ or $H$ a 0.1 Gyr old object with the same
mass out to $\sim$3000 parsecs.  Note that the $J$ and $H$ flux enhancements over black body values
for the 1 \mj$\!$ model after 0.1 Gyr are at least ten orders of magnitude.   
However, it must be remembered that
these models do not include a reflected light component from a primary.

\begin{figure}
\vspace{3.50in}
\caption{
The flux (in $\mu$Janskys) at 10 parsecs versus wavelength (in microns) from 1 \mic\ to
10 \mic\ for a 1$\!$ \mj object at ages of 0.1, 0.5, 1.0, and 5.0 Gyr.
Superposed are the positions of the $J$, $H$, $K$, and $M$ bands, 
as well as the estimated sensitivities of the three NICMOS cameras, ISO, Gemini/SOFIA,
SIRTF, and NGST.
}
\label{fig-5}
\hbox to\hsize{\hfill\includegraphics{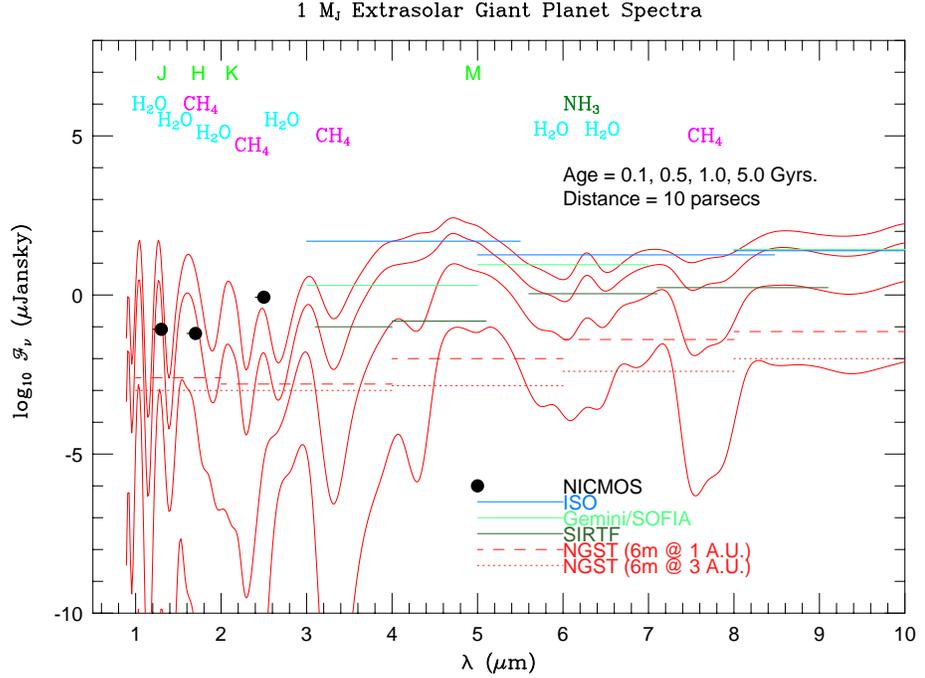}\kern+4in\hfill}
\end{figure}

\begin{figure}
\vspace{3.50in}
\caption{
The flux (in $\mu$Janskys) at 10 parsecs versus wavelength (in microns) from 1 \mic\ to
10 \mic\ for a 5$\!$ \mj object at ages of 0.1, 0.5, 1.0, and 5.0 Gyr.
Superposed are the positions of the $J$, $H$, $K$, and $M$ bands, 
as well as the estimated sensitivities of the three NICMOS cameras, ISO, Gemini/SOFIA,
SIRTF, and NGST.
}
\label{fig-6}
\hbox to\hsize{\hfill\includegraphics{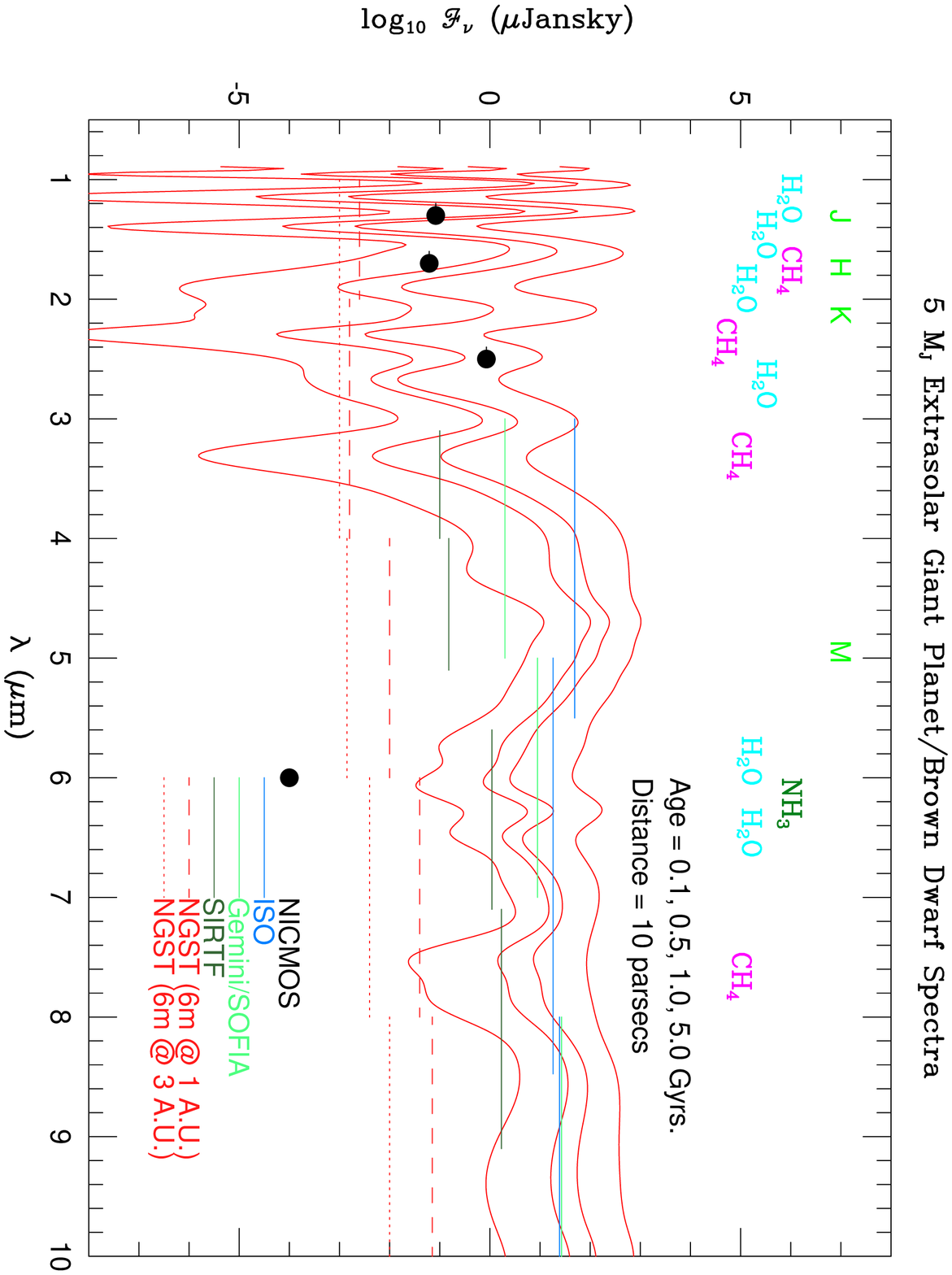}\kern+4in\hfill}
\end{figure}

\begin{figure}
\vspace{3.50in}
\caption{
The flux (in $\mu$Janskys) at 10 parsecs versus wavelength (in microns) from 1 \mic\ to
10 \mic\ for a 25$\!$ \mj object at ages of 0.1, 0.5, 1.0, and 5.0 Gyr.
Superposed are the positions of the $J$, $H$, $K$, and $M$ bands, 
as well as the estimated sensitivities of the three NICMOS cameras, ISO, Gemini/SOFIA,
SIRTF, and NGST.
}
\label{fig-7}
\hbox to\hsize{\hfill\includegraphics{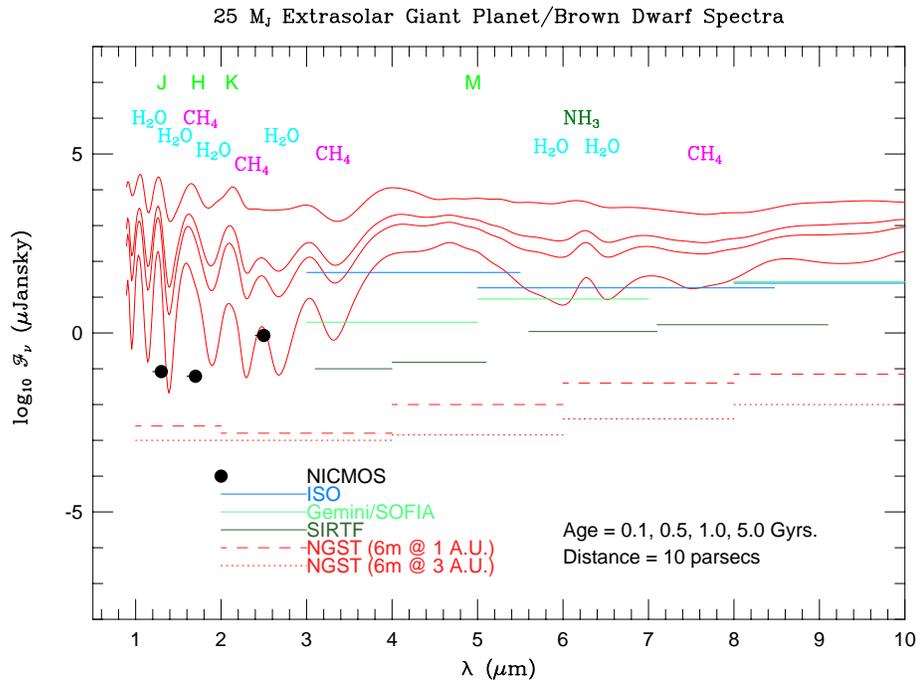}\kern+4in\hfill}
\end{figure}

\section{Infrared Colors}

To calculate IR colors, we employed the transmission curves of Bessell \& Brett (1988) and
Bessell (1990) to define the photometric bandpasses
and the model of Vega by Dreiling \& Bell (1980) for the calibration
of the magnitude scale.  
Figure 8 is a representative color--magnitude
diagram for objects with masses from 3 \mj to 40 \mj, for ages of 0.5, 1.0, and 5.0 Gyr.  
Figure 9 is a color--color diagram for the same models.   
Tables 4 \& 5 depict the infrared magnitudes and colors for various combinations of mass and age.
As Table 4 and Figure 8 suggest, the brightnesses
in the near IR are quite respectable.

\begin{figure}
\vspace{3.50in}
\caption{
Absolute $J$ vs. $J-K$ color--magnitude diagram.  Theoretical isochrones
are shown for $t$ = 0.5, 1, and 5 Gyr, along with their black body
counterparts.  The difference between black body colors and model
colors is striking.  The brown dwarf, Gliese 229B (Oppenheimer \etal\ 1995),
the young brown dwarf candidates Calar 3 and Teide 1 (Zapatero-Osorio, Rebolo, \& Martin 1997),
and late M dwarfs LHS 2924 and
GD165B (Kirkpatrick, Henry, \& Simons 1994,1995)) are plotted for comparison.
The lower main sequence is defined by a selection of M--dwarf stars from
Leggett (1992).
}
\label{fig-8}
\hbox to\hsize{\hfill\includegraphics{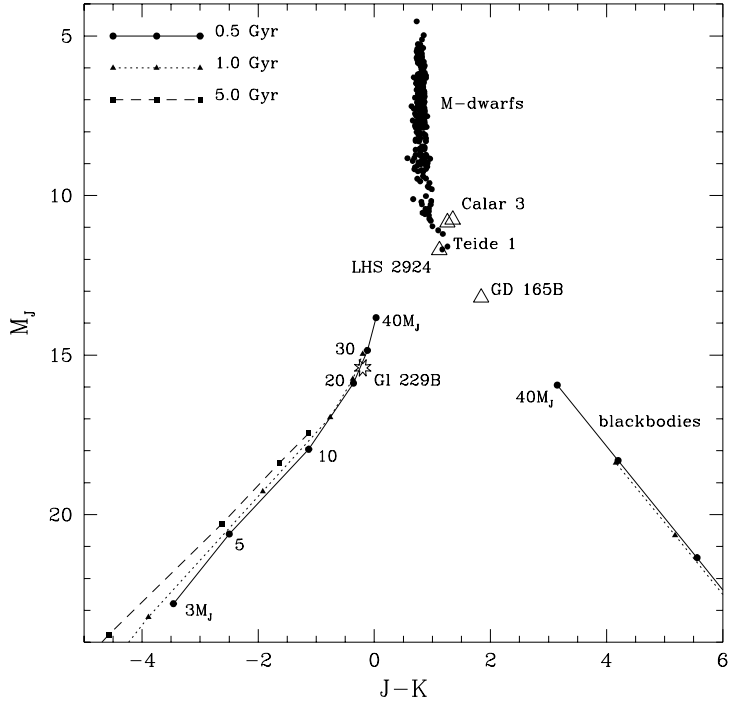}\kern+4in\hfill}
\end{figure}

\begin{figure}
\vspace{3.50in}
\caption{
$J-H$ vs. $H-K$ color--color diagram.
The edge of the main sequence as a function of metallicity, from our calculations employing Allard \& Hauschildt (1995)
atmosphere models, is shown for metallicities from
[M/H]=0 (top) to [M/H]=--3 (bottom).
Otherwise as in Figure 8.
}
\label{fig-9}
\hbox to\hsize{\hfill\includegraphics{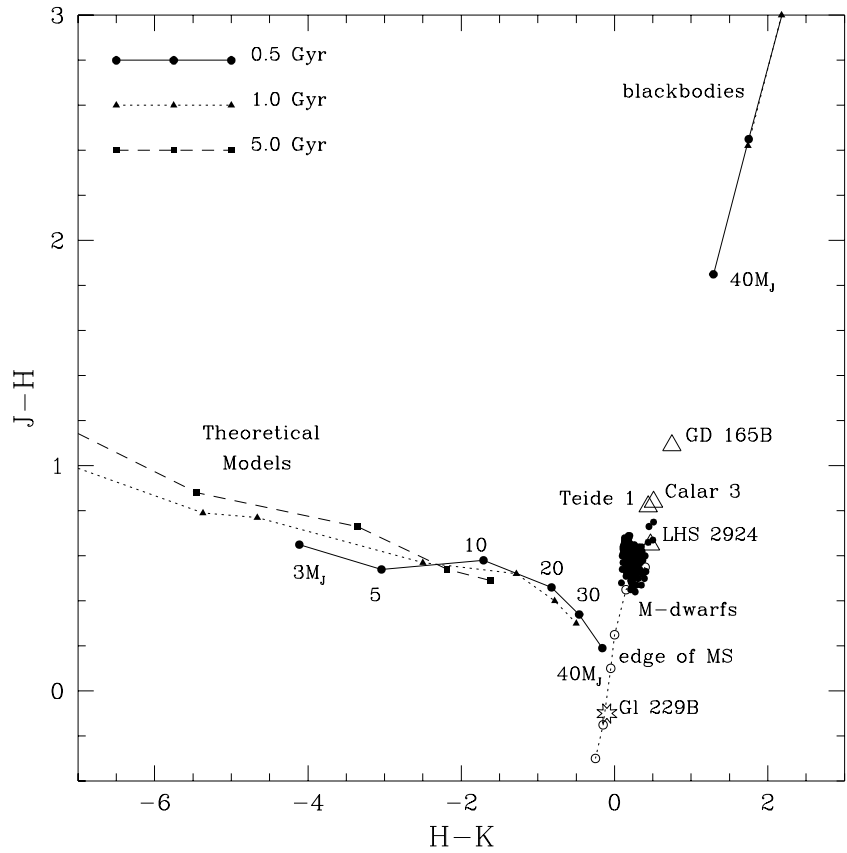}\kern+4in\hfill}
\end{figure}

For comparison, included in these figures are the corresponding black body curves, 
hot, young brown dwarf or extremely late M dwarf
candidates such
as LHS2924, GD 165B, Calar 3, and Teide 1 (Kirkpatrick, Henry, \& Simons 1994,1995; 
Zapatero-Osorio, Rebolo, \& Martin 1997),
and a sample of M dwarfs from Leggett (1992).  
These figures collectively illustrate the unique color realms occupied by extrasolar giant planets and brown dwarfs.

Figure 8 portrays the fact that the $J$ versus $J-K$ infrared H--R diagram loops back to the blue
below the edge
of the main sequence and is not a continuation of the M dwarf sequence into the red.
For $J$ versus $J-K$, the difference between the black body curves and the model curves is between 3 and 10 magnitudes.
Gl229B fits nicely on these theoretical isochrones.
The suppression of $K$ by H$_2$ and CH$_4$ features is largely responsible for this anomalous blueward
trend with decreasing mass and T$_{\rm eff}$.
The fit to Gl229B in $H$ is not as good.
This is also true of the fit to $L^{\prime}$.   Since both $H$ and  $L^{\prime}$ have
significant CH$_4$ features in them, we surmise that incompleteness 
or errors in the CH$_4$ opacity database is the cause.
$J-H$ actually reddens with decreasing T$_{\rm eff}$, but only marginally and
is still 1.5 to 4 magnitudes bluer than the corresponding black body.
That the $J-H$ and $H-K$ colors of EGPs and brown dwarfs
are many magnitudes blueward of black bodies is a firm result.

 \begin{table}
 \caption{Absolute Magnitudes of Synthetic
  BD/EGPs, [M/H]=0.0 \tablenotemark{\dag}}
 \begin{center}\scriptsize
 \begin{tabular}{ccrrrrrr}
 Mass ($M_J$) & Age (Gyr) & $M_J$ & $M_H$
  & $M_K$ & $M_{L^\prime}$ & $M_M$ & $M_N$\\
 \tableline
\\
40&  0.5&   13.83&   13.64&   13.80&   12.17&   11.86&   11.32  \\
&  1.0&   14.97&   14.67&   15.17&   13.06&   12.41&   12.25  \\
&  5.0&   17.46&   16.98&   18.59&   14.82&   13.71&   13.96  \\
\\
30&  0.5&   14.85&   14.52&   14.98&   12.94&   12.25&   12.09  \\
&  1.0&   15.75&   15.35&   16.13&   13.62&   12.73&   12.77  \\
&  5.0&   18.39&   17.85&   20.03&   15.54&   14.18&   14.46  \\
\\
20&  0.5&   15.88&   15.42&   16.24&   13.71&   12.69&   12.78  \\
&  1.0&   16.95&   16.43&   17.71&   14.51&   13.28&   13.51  \\
&  5.0&   20.28&   19.55&   22.90&   16.81&   14.95&   15.09  \\
\\
10&  0.5&   17.95&   17.37&   19.08&   15.31&   13.71&   13.97  \\
&  1.0&   19.27&   18.70&   21.19&   16.27&   14.41&   14.59  \\
&  5.0&   23.76&   22.88&   28.33&   19.22&   16.38&   15.91  \\
\\
5&  0.5&   20.61&   20.08&   23.12&   17.26&   14.95&   14.95  \\
&  1.0&   23.21&   22.43&   27.09&   18.86&   15.94&   15.62  \\
&  5.0&   29.69&   28.21&   37.20&   22.83&   18.39&   16.80  \\
\\
3&  0.5&   22.79&   22.14&   26.25&   18.64&   15.76&   15.47  \\
&  1.0&   24.82&   24.02&   29.40&   19.92&   16.58&   15.93  \\
&  5.0&   34.63&   32.10&   45.21&   25.43&   19.48&   17.50  \\
 \end{tabular}
 \end{center}
 \tablenotetext{\dag}{We employed the transmission 
 curves of Bessel \& Brett (1988) and Bessel 
 (1990) to define the photometric bandpass and 
 the model of Vega by Dreiling \& Bell (1980) 
 for the calibration of the magnitude scale.}
 \end{table}

 \begin{table}
 \caption{Color Indices of Synthetic BD/EGPs, [M/H]=0.0 \tablenotemark{\dag}}
 \begin{center}\scriptsize
 \begin{tabular}{ccrrrrr}
 Mass ($M_J$) & Age (Gyr) & $J-H$ & $J-K$ & $H-K$ & $K-L^\prime$ & $M-N$ \\
 \tableline
\\
40&  0.5&    0.19&    0.03&   -0.16&    1.63&    0.54  \\
&  1.0&    0.30&   -0.20&   -0.50&    2.11&    0.16  \\
&  5.0&    0.49&   -1.13&   -1.62&    3.78&   -0.25  \\
\\
30&  0.5&    0.34&   -0.12&   -0.46&    2.04&    0.17  \\
&  1.0&    0.40&   -0.38&   -0.78&    2.50&   -0.04  \\
&  5.0&    0.54&   -1.64&   -2.18&    4.49&   -0.28  \\
\\
20&  0.5&    0.46&   -0.36&   -0.82&    2.54&   -0.08  \\
&  1.0&    0.52&   -0.76&   -1.28&    3.20&   -0.22  \\
&  5.0&    0.73&   -2.62&   -3.35&    6.09&   -0.14  \\
\\
10&  0.5&    0.58&   -1.13&   -1.71&    3.77&   -0.26  \\
&  1.0&    0.57&   -1.93&   -2.50&    4.93&   -0.18  \\
&  5.0&    0.88&   -4.57&   -5.45&    9.12&    0.47  \\
\\
5&  0.5&    0.54&   -2.50&   -3.04&    5.86&    0.00  \\
&  1.0&    0.77&   -3.89&   -4.66&    8.24&    0.32  \\
&  5.0&    1.48&   -7.51&   -8.99&   14.38&    1.59  \\
\\
3&  0.5&    0.65&   -3.46&   -4.11&    7.61&    0.30  \\
&  1.0&    0.79&   -4.58&   -5.37&    9.47&    0.65  \\
&  5.0&    2.53&  -10.58&  -13.11&   19.78&    1.98  \\
 \end{tabular}
 \end{center}
 \tablenotetext{\dag}{We employed the transmission 
 curves of Bessel \& Brett (1988) and Bessel 
 (1990) to define the photometric bandpass and 
 the model of Vega by Dreiling \& Bell (1980) 
 for the calibration of the magnitude scale.}
 \end{table}

\section{Conclusions and Future Work}

Soon, planet and brown dwarf searches will be conducted by
NICMOS, SIRTF, Gemini/SOFIA,
ISO, NGST, LBT (Angel 1994), the MMT (Angel 1994), the VLT, Keck I \& II, COROT (transits), DENIS, 2MASS,
UKIRT, and IRTF, among other platforms.   For close companions,
advances in adaptive optics, interferometry, and coronagraphs
will be necessary.
The models we have generated of the colors and spectra of EGPs and brown
dwarfs are in aid of this quest.  
We have created a general non--gray theory of objects from 0.3 \mj to 70 \mj below $\sim$1300 K,
but the opacity of CH$_4$ and a proper treatment of silicate/iron, H$_2$O, 
and NH$_3$ clouds are future challenges
that must be met before the theory is complete. 
Since the near IR signature of nearby substellar companions will be
significantly altered by a reflected component, a theory of albedos in 
the optical and in the near IR must be developed. 
In particular, it will be useful in the future to predict the signatures
of specific systems with known orbital characteristics, primaries, and ages,
such as $\tau$ Boo, 51 Peg, $\upsilon$ And,
55 Cnc, $\rho$ CrB, 70 Vir, 16 Cyg, and 47 UMa.

It is rare that Nature conspires to make the objects of astronomical study
easier to detect than simple estimates first imply.  However,  
our calculations indicate that, whether they exist in profusion, or are merely
a minority constituent of the solar neighborhood,  
EGPs and brown dwarfs might be    
detected and characterized with a bit more ease than originally feared.

\acknowledgements 

We thank F. Allard, I. Baraffe, G. Chabrier, S. Kulkarni, J. Liebert, A. Nelson, B. Oppenheimer,  
and N. Woolf for a variety of useful contributions.  
This work was supported under NSF grants AST-9318970 and AST-9624878 and under
NASA grants NAG5-2817, NAGW-2250, and NAG2-6007.

\end{document}